\documentclass[aps,preprintnumbers,prl,twocolumn,showpacs]{revtex4}
\usepackage{amsfonts,amssymb,amsmath}            
\usepackage[]{graphics,graphicx,epsfig}            
\usepackage{amsthm}

\def\CC{{\rm\kern.24em \vrule width.04em height1.46ex depth-.07ex 
\kern-.30em C}} 
\def\P{{\rm I\kern-.25em P}} 
\def\RR{{\rm 
         \vrule width.04em height1.58ex depth-.0ex 
         \kern-.04em R}} 
 
\def\RR{{\rm\kern.24em \vrule width.04em height1.46ex depth-.07ex 
\kern-.30em R}} 
\def\P{{\rm I\kern-.25em P}} 
\def\RR{{\rm 
         \vrule width.04em height1.58ex depth-.0ex 
         \kern-.04em R}}

\newcommand{\be}{\begin{equation}} 
\newcommand{\ee}{\end{equation}} 
\newcommand{\bq}{\begin{eqnarray}} 
\newcommand{\eq}{\end{eqnarray}}

\newcommand{\sx}{\sigma^x} 
\newcommand{\sy}{\sigma^y} 
\newcommand{\sz}{\sigma^z} 
\newcommand{\ket}[1]{\left | #1\right \rangle} 
\newcommand{\bra}[1]{\left \langle #1\right |} 
\date{\today} 
 
\begin{document} 
\title{Geometric phases and criticality in spin chain systems}
\author{Angelo C. M. Carollo and Jiannis K. Pachos }
\address{
Centre for Quantum Computation, Department of Applied Mathematics and
Theoretical Physics, \\ University
of Cambridge, Wilberforce Road, Cambridge CB3 0WA, UK }

\begin{abstract}
A relation between geometric phases and criticality
of spin chains is established. As a result, we show how geometric phases can be
exploited as a tool to detect regions of criticality without having
to undergo a quantum phase transition. We analytically evaluate the
geometric phase that correspond to the ground and excited states of
the anisotropic XY model in the presence of a transverse magnetic
field when the direction of the anisotropy is adiabatically
rotated. It is demonstrated that the resulting phase is resilient
against the main sources of errors. A physical realization with ultra-cold atoms in optical lattices
is presented.
\end{abstract}


\maketitle

\vspace{0.2cm}

Since the discovery by Berry~\cite{Berry84}, geometric phases
in quantum mechanics have been the subject of a variety of theoretical
and experimental investigations~\cite{ShapereW89}. Possible
applications range from optics and molecular physics to fundamental
quantum mechanics and quantum computation~\cite{Bohm}. In condensed
matter physics a variety of phenomena have been understood as a
manifestation of topological or geometric
phases~\cite{Korepin,Niu,Schutz,Bruno,Wiemer}. An interesting open question
is whether the geometric phases can be used to investigate the physics and
the behavior of condensed matter systems. Here we show how to exploit
the geometric phase as an essential tool to reveal quantum critical phenomena in many-body
quantum systems. Indeed, quantum phase transitions are accompanied by
a qualitative change in the nature of classical correlations and their
description is clearly one of the major interests in condensed matter
physics~\cite{Thouless98,Sachdev}. Such drastic changes in the
properties of ground states are often due to the presence of points of degeneracy
 and are reflected in the geometry of the Hilbert space of the
system. The geometric phase, which is a measure of the curvature of the
Hilbert space, is able to capture them, thereby revealing
critical behavior. This provides the means to detect, not only
theoretically, but also experimentally the presence of criticality
without having to undergo a quantum phase transition.

In this letter we analyze the XY spin chain model and the geometric phase
that corresponds to the XX criticality. Since the XY model is exactly
solvable and still presents a rich structure it offers a benchmark to test
the properties of geometric phases in the proximity of a quantum phase
transition. Indeed, we observe that, an excitation of the model obtains a
non-trivial Berry phase if and only if it circulates a region of
criticality. The generation of this phase can be traced down to the presence
of a conical intersection of the energy levels located at the XX
criticality. This geometric interpretation reveals a relation between the
critical exponents of the model. The insights provided here shed light
into the understanding of more general systems, where analytic solutions
might not be available. A physical implementation is proposed with
ultra-cold atoms superposed by optical
lattices~\cite{Pachos2,Pachos1}. It utilizes Raman activated tunneling
transitions as well as coherent drive of the system via
Bragg scattering. The independence of the generated phase from the
number of atoms, its topological nature and its resilience against
control errors makes the proposal appealing for experimental realization.

\begin{figure}
\begin{center}
\includegraphics[width=3.3in]{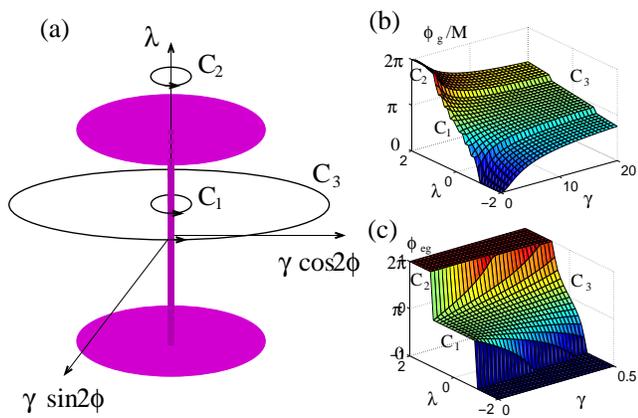}
\vspace{0.5cm}
\caption{(a) The regions of criticality are presented as a function
of the Hamiltonian parameters $ \lambda $, $\gamma
\cos{2 \phi}$ and $\gamma \sin{2 \phi}$. The geometric
phase corresponding to the ground state (b) and the relative one
between the ground and excited state (c) as a function of the path
parameters $\lambda $ and $\gamma$. Values of the geometric phase
corresponding to the loops $C_1$, $C_2$ and $C_3$ in (a) are also
indicated.} 
\label{caruli}
\end{center}
\end{figure}

Consider the one dimensional spin-1/2 XY model, with N spins, in a
transverse magnetic field. This chain has nearest neighbor interactions with Hamiltonian given by
\begin{equation}
\label{HXYModel}
H = -\sum_{l=-M}^{M} \left(\frac{1 + \gamma}{2}\sx_l
\sx_{l+1} + \frac{1 - \gamma}{2}\sy_l \sy_{l+1}
+\frac{\lambda}{2}\sz_l \right),\nonumber
\end{equation}
where $M=(N-1)/2$ for $N$ odd. In particular, we are interested in the
Hamiltonian that can be obtained by applying a rotation
of $\phi$, around the z-direction, to each spin
\begin{equation}
 \label{HXYphi}
H(\phi) = g(\phi)H g^\dag (\phi)\quad \text{ with }\quad g(\phi) =
\prod_{l=-M}^{M} e^{i\sz_l\phi/2}.
\end{equation}
The family of Hamiltonians that is parameterized by $\phi$ is clearly
isospectral and, therefore, the critical behavior is independent from $\phi$. This is reflected in the symmetric structure of the regions of
criticality shown in Figure~\ref{caruli}. In addition, due to its bilinear
form, $H(\phi)$ is $\pi$-periodic in $\phi$. The Hamiltonian $H(\phi)$ can
be diagonalized by a standard procedure, which can be summarized in  the
following three steps: (i) the Jordan-Wigner transformation, which converts
the spin operators into fermionic operators via the relations,
$a_l=(\prod_{m<l}\sz_m)(\sx_l+i\sy_l)/2$; (ii) their Fourier transformation,
$d_k=\frac{1}{\sqrt{N}}\sum_{l} a_l e^{-i2\pi l k/N}$, with $k=-M,\dots, M$;
and (iii) the Bogoliubov transformation, which defines the fermionic
operators, $b_k=d_k \cos\frac{\theta_k}{2}-id_{-k}^\dag
e^{2i\phi}\sin\frac{\theta_k}{2}$, where the angle $\theta_k$ is defined by
$\cos \theta_k=
\epsilon_k/\Lambda_k$ with
$\epsilon_k=\cos{2\pi k \over N}- \lambda$ and
$\Lambda_k=\sqrt{\epsilon_k^2+
\gamma^2 \sin^2{2\pi k \over N}}$.
These procedures diagonalize the Hamiltonian to a form
\begin{equation}
\label{Hdiagonal}
H(\phi)=\sum_{k=-M}^{M}\Lambda_k b_k^\dag b_k.
\end{equation}
The ground state of $H(\phi)$ is the vacuum
$\ket{g}$ of the fermionic modes, $b_k$,
given by
\begin{equation}\label{groundstate}
\ket{g}\!=\!\!\!\prod_{\otimes k>0}\!\!\Big(\!\cos {\theta_k \over 2}
\ket{0}_{\!k}\ket{0}_{\!\!-k}
\!\!-ie^{2i\phi} \sin {\theta_k \over 2}
\ket{1}_{\!k} \ket{1}_{\!\!-k}\!\Big),
\end{equation}
where $\ket{0}_{k}$ and $\ket{1}_k$ are the
vacuum and single excitation of the k-th mode,
$d_k$, respectively.
The regions of criticality which appear when the ground and first excited
states become degenerate are shown for this model in Figure
\ref{caruli}(a). The XX model, which corresponds to $\gamma=0$, has criticality region
along the line between $ \lambda = 1 $ and $\lambda=-1$. The region of the Ising
model phase transition are the two planes at $\lambda =1 $ and
$\lambda=-1$. The interesting paths of evolution
for generating a Berry phase are those for which the state of the
system can evolve around a region of criticality. Here we shall focus
on the criticality region corresponding to the XX
model \cite{Sachdev}, which can be encircled by adiabatically varying
the angle $\phi$ from $0$ to $\pi$. The corresponding Berry phase
of ground and first excited states can be evaluated as a function of $\lambda$ and $\gamma$.

Using the standard formula it is easy to show that the Berry phase of
the ground state $\ket{g}$ is given by
\begin{equation}
\label{geomphase}
\varphi_g = -i\int_0^\pi\bra{g}\frac{\partial}{\partial\phi}
\ket{g} = \sum_{k>0}\pi(1-\cos\theta_k).
\end{equation}
This result can be understood by considering the form of $\ket{g}$, which is
a tensor product of states, each lying in the two dimensional Hilbert space
spanned by $\ket{0}_{k}\ket{0}_{-k}$ and $\ket{1}_k\ket{1}_{-k}$. For each
value of $k(>0)$, the state in each of these two-dimensional Hilbert spaces
can be represented as a Bloch vector with coordinates $(2\phi,\theta_k)$. A
change in the parameter $\phi$ determines a rotation of each Bloch vector
about the $z$ direction. A closed circle will, therefore, produce an overall
phase given by the sum of the individual phases as given in
(\ref{geomphase}) and illustrated in Figure \ref{caruli}(b). It is worth
noticing that when $\gamma$ is small, the overall phase is strongly
dependent on the exact values of $\lambda$ and $\gamma$ due to the (large)
proportionality factor $M$. On the other hand, when $\gamma \gg \lambda$,
each pair of $k$ and $-k$ states from (\ref{groundstate}) contributes almost
equally to the overall Berry phase (see Figure \ref{caruli}(b)). In this
case, we obtain $\varphi_g/M \approx \pi$ which makes effectively the
dependence of the overall Berry phase on the parameters $ \gamma $,
$\lambda$ much weaker.

Of particular interest is the relative geometric phase between the
first excited and ground states given by the difference of the Berry phases acquired
by these two states. The first excited state is given by $\ket{e_{k_0}}
\equiv b^\dag_{k_0}\ket{g}$, i.e. it is the single excitation of the $b$
fermionic field, with $k_0$ corresponding to the minimum value of the energy
$\Lambda_k$. The relative geometric phase then becomes
\begin{equation}
\label{connectionExcited}
\varphi_{eg} \equiv \varphi_e-\varphi_g =
-i \oint \bra{g} b_k \frac{\partial b^\dag_k}{\partial
 \phi}\ket{g}.
\end{equation}
If the parameter $\phi$ is adiabatically changed from $0$ to $\pi$, we
obtain $\varphi_{eg}= -\pi(1-\cos\theta_{k_0})$ which, in the
thermodynamical limit ($N \to \infty$), takes the form
\begin{equation}
\label{GPExcGrndSmallGamma}
\varphi_{eg}=\left\{\begin{array}{cl}
0,      &   \text{for $|\lambda |>1-\gamma^2$} \\
-\pi+{\pi \lambda \gamma \over
  \sqrt{(1-\gamma^2)(1-\gamma^2-\lambda^2)}},     &    \text{for
  $|\lambda |<1-\gamma^2$}.
\end{array}\right.
\end{equation}
As can be seen from Figure~\ref{caruli}(c), the most interesting behavior of
$\varphi_{eg}$ is obtained in the case of $\gamma$ small compared to
$\lambda$. In this case $\varphi_{eg}$ behaves as a step function, giving
either $\pi$ or $0$ phase, depending on whether $|\lambda|<1$ or
$|\lambda|>1$, respectively. This behavior is precisely determined from
whether the corresponding loop encloses a critical point or not and can be
used as a witness of its presence. In particular, in the $|\lambda
|<1-\gamma^2$ case the first term corresponds to a purely topological phase,
while the second is a geometric contribution \cite{Bohm}.

This is related to a more general property: a non-trivial Berry
phase is generated when the closed loop in the parameter space is
spanned in the vicinity of a degeneracy point. The topological nature of this phase is evident in the case of a loop contracting to a
point. If, in this limit, the Berry phase remains non-vanishing, then
the contraction point identifies the position of the degeneracy.
As a critical point is also a point of degeneracy it is apparent that
this procedure can be used to locate the criticality of an Hamiltonian.

To better understand the properties of the relative geometric phase in our model,
let's focus on the region of parameters with
$\gamma\ll 1$, i.e. for a loop around the region of criticality which contracting to a point. 
In this case, it can be shown that the Hamiltonian, when
restricted to its lowest energy modes $d_{k_0}$ and $d_{-k_0}$, can be
casted in a {\em real} form and, for $|\lambda|<1$, its eigenvalues present
a {\em conical intersection} centered at $\gamma=0$. It is well known that
when a closed path is taken within the real domain of a Hamiltonian, a
topological phase shift $\pi$ occurs only when a conical intersection is
enclosed. In the present case, the conical intersection corresponds to a
point of degeneracy where the XX criticality occurs and it is revealed by
the topological term in the relative geometric phase $\varphi_{eg}$.
 It is
worth noticing that the presence of a conical intersection indicates that
the energy gap scales linearly with respect to the coupling $\gamma$
when approaching the degeneracy point. This implies that the
critical exponents of the energy, $z$, and of the correlation length, $\nu$,
satisfy the relation $z\nu=1$ which is indeed the case for the XX
criticality~\cite{Sachdev}.


From an experimental point of view, this method of obtaining a geometric phase is
robust against relevant sources of errors. The most significant one
originates from undesired transitions induced by the time dependence of the
Hamiltonian. In any adiabatic process the occurrence of these transitions
is hindered primarily by the presence of a finite energy separation between
neighboring levels. However, as in many condensed matter systems, these
energy separations may decrease quite rapidly as the thermodynamical limit is
approached. In fact, in this model, the gaps between the energies of the lowest
excited states scale typically as $N^{-2}$, which might substantially
reduce the applicability of the adiabatic approximation in a truly many-body
setup. However, in the adiabatic evolution proposed here, all the
transitions between closely separated energy levels are prohibited due to
symmetry reasons. It is apparent form equation~(\ref{HXYphi}) that the
adiabatic evolution is generated by the Hamiltonian term $\tilde{H}\propto
\sum_k{\sigma_k^z}=\sum_k d_k^\dag d_k$, which preserves the excitations in
each fermionic mode, $d_k$. In fact, the excitations
$\ket{e_{k}}=b_k^\dag\ket{g}$ of model~(\ref{HXYModel}) are also
eigenstates of $\tilde{H}$. Therefore, the only allowed transitions
induced by $\tilde{H}$ are
between the ground state and the doubly excited states $b_k^\dag
b_{-k}^\dag\ket{g}$. The corresponding finite energy gap $2\Lambda_k$ is
then sufficient to adiabatically prevent these unwanted transitions, even in
the thermodynamical limit. Moreover, this symmetry constraint guarantees the
viability of the adiabatic approximation also when energy level
crossings occur between singly excited states $\ket{e_k}$'s.

Another important feature of this model is the behavior of the phase under
errors in the initial preparation. The smooth dependence of the
geometric phase on the momentum $\chi=2\pi k/N$, insures that a small
deviation $\delta=\chi-\chi_0$ from the value $\chi_0=2\pi k_0/N$
(corresponding to first excited state, i.e. the minimum of the function
$\Lambda(\chi)$) does not drastically affect the interference pattern.
Indeed a perturbation analysis shows that states with a small momentum
difference, $\delta$, from $\chi_0$ acquire
geometric phases with values shifted by an amount proportionally to $\delta$. In
particular, due to the independence of the function $\Lambda(\chi)$ from the
system size, $N$, the spread of the geometric phase only depends on
$\delta$ and remains unchanged in the thermodynamical
limit. Naturally, this effect is accompanied by a
reduction of the visibility and a shift in the value of the measured phase,
which, however, only depend on the preparation and measurement accuracies
and does not depend on the system size.


This construction, apart from its theoretical interest, offers a possible
experimental method to detect critical regions without the need to cross
them, which undermines the ability to identify the state of the system due
to the presence of degeneracy. In particular, we shall implement this model
with optical lattices. To this end, consider two bosonic species labelled by
$\sigma=a, b$ that can be given by two hyperfine levels of an atom. Each one
can be trapped by employing two in-phase one dimensional optical lattices.
The tunneling of atoms between neighboring sites is described by
$V=-\sum_{l\sigma} (J_\sigma a_{l\sigma}^\dagger a_{(l+1)
\sigma} +\text{H.c.})$. When two or more atoms are present in the same site,
they experience collisions given by $H^{(0)}=
\sum_{l\sigma \sigma'}  {U_{\sigma \sigma'} \over 2}
a^\dagger_{l\sigma} a^\dagger_{l\sigma'} a_{l\sigma'} a_{l\sigma}$. We
shall consider the limit $J \ll U$ where the system is in the Mott
insulator regime \cite{Raithel} with one atom per lattice site. In this regime,
the effective evolution is obtained by adiabatic elimination of the
states with a population of two or more atoms per site, which are
energetically unfavorable. Hence, to describe the Hilbert space of
interest, we can employ the pseudospin basis of $|\!\!\uparrow\rangle
\equiv |n_l^a=1,n_l^b=0\rangle$ 
and $|\!\!\downarrow\rangle\equiv |n_l^a=0,n_l^b=1\rangle $, for lattice site $l$, and the
effective evolution can be expressed in terms of the corresponding Pauli operators.
It is easily verified that when the tunneling coupling of both atomic
species is activated, the following exchange interaction is realized
between neighboring sites~ \cite{Pachos1,Kuklov,Duan}, 
\be
H_1=-\frac{J_a J_b}{U_{ab}} \sum_l \left(
\sigma^x_l\sigma^x_{l+1}+\sigma^y_l\sigma^y_{l+1}\right).
\label{Ham1}
\ee
In order to create an anisotropy between the x and y spin directions,
we activate a tunneling by means of Raman couplings
\cite{Pachos1,Duan}. Application of two standing lasers
$L_1$ and $L_2$, with zeros of their intensities
at the lattice sites and with phase difference $\phi$, can induce
tunneling of the state $|+\rangle \equiv ({e 
  ^{-i\phi /2} |a\rangle + e ^{i\phi /2} |b\rangle) /  \sqrt{2}}$.
The resulting tunneling term is given by
$V_c=J_c \sum_l (c^\dagger_l c^{}_{l+1} +\text{H.c.})$,
where $c_l$ is the annihilation operator of $|+\rangle$ state
particles. The tunneling coupling, $J_c $, is given by
the potential barrier of the initial optical lattice superposed by the
potential reduction due to the Raman transition.
The resulting evolution is dominated by an effective Hamiltonian
given, up to a readily compensated Zeeman term, by
\be
H_2=-{1 \over 2}{J_c^2 \over U_{ab}} \sum_l
g(\phi)\sigma^x_l\sigma^x_{l+1}g^\dagger(\phi),
\label{Ham0}
\ee
where $g(\phi)$ was defined in (\ref{HXYphi}).
Combining the rotationally invariant Heisenberg interaction $H_1$ with $H_2$
gives the rotated XY Hamiltonian described by equation~(\ref{HXYphi}),
where the parameter $\gamma$ is given by $J_c^2/(2\epsilon U_{ab})$ and
$\epsilon = (2J_a J_b +J_c^2/2)/U_{ab}$ is the overall energy scale
multiplying the Hamiltonian~(\ref{HXYModel}).

At this point we would like to generate and measure the relative phase between the
ground and excited states. To create the first spin chain excitation,
$\ket{e_{k_0}}$, in the optical lattice setup we employ two-photon
Bragg spectroscopy \cite{Stenger}. The
use of a two-photon process allows for an accurate energy
resolution of the state we want to excite, which is crucial
considering that the energy gap, $\Lambda_{k_0}$, is  small compared
to the tunneling and collisional 
couplings \cite{comment}. For example, an amplitude modulation of the axial
lattice potential, $V_x(x,t) = [V_{x,0} + A_{\text{mod}} \sin (2 \pi
\nu_{\text{mod}}t)] \sin^2(kx)$, can be employed \cite{Stoferle} with a
frequency that satisfies the resonant Bragg condition, $\nu_{\text{mod}} =
\Lambda_{k_0}$. This will activate a coherent energy population
transfer from the ground state to the excited one with a transition
amplitude proportional to the modulation amplitude, $A_{\text{mod}}$. Such
an experimental technique has already been employed to study the atom-number
excitations in an optical lattice near its criticality between the
superfluid and the Mott insulator regime \cite{Stoferle}. To detect the
Berry phase of the excited state, one can initially create the superposition
$\ket{g}+\ket{e_{k_0}}$ with an appropriately timed lattice modulation with
frequency $\nu_{\text{mod}}=\Lambda_{k_0}$. The Berry phase procedure is
then applied to generate a phase difference of $\pi$ between the ground and
the excited states giving $\ket{g}-\ket{e_{k_0}}$. Applying the same
modulation procedure as before will finally result in the state
$\ket{e_{k_0}}$. To distinguish between the states $\ket{g}$ and
$\ket{e_{k_0}}$ one can measure the corresponding spin-spin
correlators, e.g. by atomic scattering of fast atoms off the lattice
\cite{Kuklov2}.

To conclude, we have presented a method that theoretically, as well as
experimentally, enables the detection of regions of criticality
through the geometric phase, without the need for the system to
experience phase transitions. The latter is experimentally hard to
realize as the adiabaticity condition breaks down and the state of the
system is no longer faithfully represented by the ground state. The
origin of the geometric phase can be ascribed to the existence of
degeneracy points in the parameter space of the Hamiltonian. Hence, a
criticality point can be detected by performing a looping trajectory
around it and detecting whether or not a non-zero Berry phase has been
generated. For the case of the XY model the topological nature of the
resulting phase pinned to the value, $\varphi_{eg} \approx \pi$, is
revealed by its resilience with respect
to small deformations of the loop. Topological phases are
inherently resilient against control errors, a property that can be
proved to be of a great advantage when considering many-body
systems. Such a study can be theoretically performed on any system
which can be analytically elaborated such as the case of the cluster
Hamiltonian~\cite{Pachos2}, or exploited numerically when analytic
solutions are not known. 
The generalization of these results to a wide variety of critical
phenomena and their relation to the critical exponents is a
promising and challenging question which deserves extensive future
investigation.

{\em Acknowledgements}. We would like to thank E. Rico for inspiring
discussions. This work was supported by the EU TopQIP network and the Royal
Society.



\begin{references}
\bibitem{comment} The energy gap, $\Lambda_{k_0}$, is of the order of
$\lambda,\gamma$. In the optical lattice implementation $\lambda$ is
manipulated externaly by a heavily detuned laser fields, while $\gamma$ is of
the order of $J_c^2/(J_a J_b)$.
\end{references}
\end{document}